\documentstyle[12pt,aasms4]{article}
\slugcomment{In Press: The Astrophysical Journal}
\lefthead{Nemiroff}
\righthead{Brightest Microlensing Events}

\begin{document}

\title{Towards Locating the Brightest Microlensing Events on the Sky}

\author{Robert J. Nemiroff}
\affil{Department of Physics, Michigan 
Technological University,
Houghton, MI  49931}

\begin{abstract}
It is estimated that a star brighter than visual magnitude 17 is undergoing a detectable gravitational microlensing event, somewhere on the sky, at any given time.  It is assumed that both lenses and sources are normal stars drawn from a standard Bahcall-Soneira model of our Galaxy.  Furthermore, over the time scale of a year, a star 15th magnitude or brighter should undergo a detectable gravitational lens amplification.  Detecting and studying the microlensing event rate among the brightest 10$^8$ stars could yield a better understanding of Galactic stellar and dark matter distributions.  Diligent tracking of bright microlensing events with even small telescopes might detect planets orbiting these stellar lenses.
\end{abstract}


\section{ Introduction }

The ability of one star to gravitationally amplify the light from another was considered previously by Einstein (1936), and developed more generally by, for example, Refsdal (1964).  Realistic rate estimates for gravitational microlensing by massive parallel photometry were made by Paczynski (1986) for the LMC and again by Paczynski (1991) for the Galactic bulge.

The first measurements of microlensing were reported by
Udalski et al. (1993) and Alcock et al. (1993).  
Several microlensing searches are currently underway, including the OGLE collaboration (Udalski et al. 1993), the MACHO collaboration (Alcock et al. 1993), the EROS collaboration (Aubourg et al. 1993), the AGAPE collaboration (Alard et al. 1995), and a collaboration monitoring M31 (Crotts et al. 1992).  Groups that track microlensing events, once announced, include the PLANET collaboration (Albrow et al. 1995), and the GMAN collaboration (Becker et al. 1997).

Combined, microlensing search collaborations currently monitor tens of millions of stars down to a visual magnitude of about 19.5 over a small fraction of the sky, currently reporting over 100 microlensing events (for a good review, see Paczynski 1996).  Of the first 45 gravitational lens candidate events toward the Galactic bulge, the brightest star recorded to undergo a microlensing event by the MACHO collaboration was of unlensed visual magnitude 16.4 (Alcock et al. 1997).

A primary motivation for this study was curiosity about how bright a star was undergoing a detectable microlensing event at any given time.  Previously, Colley \& Gott (1995) estimated that a microlensing events becomes bright enough for naked eye observation nearly every 5000 years.  Perhaps, however, a star that is significantly brighter than those being monitored is {\it currently} undergoing detectable microlensing.  Microlensing of this star might be relatively easy to track, as it would be relatively bright and in a less crowded field.  This paper is an attempt to estimate the magnitudes of the brightest microlensing events on the entire sky, and to discuss the possible advantages of finding these events. \S 2 will detail how this estimate will be made. In \S 3, results will be presented, while some discussion and conclusions will be given in \S 4.

\section{ All-Sky Microlensing Rate Estimates }

Nemiroff (1986) gave a first estimate of the all-sky microlensing rate.  There it was shown that to amplify source light at least $A_{abs}$, a lens with Schwarzschild radius $R_S$ must fall into a ``detection volume" $V = (2/3) \pi R_S D_{OS}^2 \Phi_{abs}$, where distance $D$ has subscripts $O$, $L$, and $S$ which stand for observer, lens, and source, respectively.  $\Phi_{abs} = A/ \sqrt{A^2 - 1} - 1$, and here carries the subscript ``abs" which refers to absolute amount of source magnification.

Assuming all stars are identical and uniformly distributed, the likely number of lens events $N$ above absolute amplitude $A_{abs}$ out to distance $D_{OS}$ is found by summing the space internal to all the lens detection volumes, so that $N = (8/15) \pi^2 n^2 \Phi_{abs} R_S D_{OS}^5$, where $n$ is the number density of stars.
This early estimate did not include the brightness of the lensing star itself. Estimates of the effects of the brightness of the lens or an unlensed component have been done previously with regard to existing microlensing surveys by Kamionkowski (1995), Nemiroff (1997) and Han (1998). Nemiroff (1997) has shown that for a field of lenses with identical brightness, the apparent optical depth is a factor of about six less than the absolute optical depth, which includes the assumption that the lens is dark.  Including this factor in the previous estimate, the number of lenses expected above the {\it apparent} amplification factor of $A_{app}$ would be
 \begin{equation}
  N \sim (4/45) \pi^2 n^2 \Phi_{app} R_S D_{OS}^5 .
 \end{equation}

Given $n =$ 0.1 stars pc$^{-3}$ and $R_S = 3$ km, 
$\Phi_{app} = 1/2$, the single closest lensing event is expected at $D_{min} = 1200$ pc.  The total number of stars in this volume is 
$N ({\rm stars}) = (4/3) \pi D_{OS}^3 n$ = 7.0 x 10$^8$.  
Given the canonical star is like the Sun and has absolute visual magnitude $M_V \sim +5$, its apparent magnitude at 1200 pc is an estimate of the brightest microlensing event occurring on the sky at any one time: $m_V \sim 15.4$.

Although instructive in calibrating intuition, this estimate did not include the anisotropic and inhomogeneous stellar distributions that surround our Earth and our Galaxy. A more accurate estimate might be significantly different.  
Toward this end, an all-sky microlensing rate estimate is made here by employing the Galactic model published originally by Bahcall \& Soneira (1980) and subsequently updated by Bahcall (1986). The Bahcall-Soneira Galactic model obtains a reasonable fit to the observed star counts with a relatively simple disk and spheroid model.  Bahcall (1986) demonstrates accuracy of the model in several directions down past 20th magnitude, although all of these directions are at least 20 degrees away from the Galactic plane.  The model also included large-scale affects of reddening and obscuration by Galactic dust. 

A FORTRAN-program version of the Bahcall-Soneira model was downloaded from J. Bahcall's web site in December 1997 and adapted to address microlensing.  
The Bahcall-Soneira code tracks both main-sequence and giant stars for both a Galactic disk and spheroid.  For main-sequence disk stars, the code was run with
absolute visual magnitude $M_V$ cut-offs of $-6$ at the bright end at $+16.5$ at the dim end.  For disk giants, $M_V$ was taken to be $-1.5$ at the bright end.  For spheroid stars, $M_V$ was taken to be $-3$ at the bright end.

In general, the total number of lensing events visible is given by
 \begin{equation} 
   N =            \int_0^{4 \pi} \ d\Omega
                  \int_0^{D_{OS}^{MAX}} D_{OS}^2 \ dD_{OS}
                  \int_{M_{DIM}}^{M_{BRT}} n(M) \ dM_S
                  \int_0^{D_{OS}} dD_{OL}
                  \int_{M_{DIM}}^{M_{BRT}} n(M) \ 
                  \pi b_L^2 \ dM_L ,
 \end{equation}
where $\Omega$ is a solid angle on the sky, $D_{OS}^{MAX}$ is the maximum source distance, $M_{DIM}$ is the cutoff for absolute visual magnitude at the dim end, $M_{BRT}$ is the cutoff for $M_V$ at the bright end, $M_S$ is the absolute visual magnitude of a source star, and $M_L$ is the absolute visual magnitude for the lens.  The impact parameter $b_L$ is the furthest distance from the observer-source axis a lens could be placed, at distance $D_{OL}$, and still amplify source light by the amount $A_{app}$ needed for detection.  From Nemiroff (1986, 1989, 1997) this impact parameter was found to be
 \begin{equation}
  b_L = \sqrt{ 4 R_S D_{OL} D_{LS} \Phi_{app} \over D_{OS} } .
 \end{equation}
The brightness of the stellar lens itself works to dampen the magnitude of observed microlensing events.  More precisely, to achieve an apparent lensing amplification of $A_{app}$, the lens must create an absolute amplification
$A_{abs} > A_{app}$ to be seen over its own brightness (Nemiroff 1997): $A_{abs} = A_{app} (1 + l_L/l_S) - l_L/l_S$, where $l$ indicates apparent brightness.

Microlensing rates from several candidate scenarios were made.  In the first, it was assumed that all stars brighter than a given unlensed magnitude limit would be monitored for microlensing, a case referred to as ``without magnification bias."  In the second scenario, it was assumed that a star more faint than a given magnitude limit could be amplified above the limit and be detected.  The second scenario therefore includes magnification bias (Nemiroff 1994).  In the third scenario, only stars in the Galactic disk of the Bahcall-Soneira model were included, without magnification bias, to give a indication of the angular distribution of the microlensing events.

In all of the computations, giant stars were not allowed to be lenses, since one or both amplified images might impact the lens surface and be absorbed (Ftaclas 1998).  The mass of main sequence stellar lenses was taken from Mihalas \& Binney (1981) to be $M / M_{\odot} = (L / L_{\odot})^{0.3125}$.

Equation 2 above was used to compute $N(m)$, the number of expected lens events anywhere on the sky.  For any magnitude limit $m$, however, the maximum expected amplification $A_{Max}$ on the sky can be computed directly from $N(m)$ since $N(m)$ scales as $\Phi_{app}$.  In fact, when $\Phi_{app} = 1/2$, $N = \tau$, the optical depth (Vietri \& Ostriker 1983; Nemiroff 1989).  When $N$ events are expected at the level of $\Phi_{app} = 1/2$ (where $A_{app} =$ 1.34), then fewer events are expected at a higher $A_{app}$ (Nemiroff 1997).  Specifically, the level of $\Phi_{app}$ where one expects a single event is at $\Phi_{app} / (1/2) = N$.  The corresponding amplitude is  
 \begin{equation}
  A = { {N + 2} \over \sqrt{N^2 + 4N} } .
 \end{equation}

\section{ Results }

Figure 1 shows a plot of the expected optical depth to stellar microlensing as a function of apparent visual magnitude $m_V$. When the inverse of the optical depth becomes on the order of the number of stars being observed, then there is a 63 \% chance that at least one of the stars is amplified by a factor greater than 1.34, and detection of gravitational lensing becomes likely.  Including stars in both the Galactic disk and the central spheroid, stellar microlensing becomes likely between 16th and 17th magnitude. The Bahcall-Soneira model used predicts almost exactly 100 million stars exist over the entire sky that are brighter than 17th magnitude.

Perhaps unexpected is that Figure 1 shows that optical depth can actually be a decreasing function of the survey apparent magnitude limit.  This is caused by the shape of the stellar luminosity function, which may cause a disproportionate number of stars in a given magnitude range to be relatively close to the observer, hence contributing relatively little to the optical depth. This decrease in optical depth means only that the probability of detectable lensing {\it per star} has decreased.  Since star counts increase greatly as magnitude dims, the total probability of finding microlensing somewhere on the sky at fainter magnitudes still sharply increases.

Figure 2 shows plots of the expected maximum amplitude of the brightest lens event on the sky as a function of the faintest apparent visual magnitude monitored.  Inspection of this plot again indicates at about 17th magnitude, an $A_{Max} > 1.34$ event is expected, even without magnification bias.  A search sensitive to magnification bias could discover that a star temporarily brighter than 16th magnitude is actually being detectably microlensed, somewhere on the sky.  The more stable the relative photometry, the smaller the microlensing enhancements it would be sensitive to, the brighter the star that would show detectable microlensing.  Alternatively, were all stars brighter than 20th magnitude monitored, at least one would be expected to be undergoing an $A > 3.2$ event at any given time.

Figure 3 shows the number of $A > 1.34$ lensing events expected over the sky as a function of visual magnitude.  This plot once again indicates a 63 \% chance that one star with unlensed magnitude brighter than 17 is being amplified by $A > 1.34$, and that it is likely a star appearing to be brighter than 16th magnitude is undergoing an $A > 1.34$ microlensing event.  

Additionally, Figure 3 indicates that on average, two stars with unlensed magnitude brighter than 18 are undergoing detectable microlensing, while potentially 9 stars brighter than 20th magnitude are expected to undergo simultaneously detectable gravitational lens magnifications with $A > 1.34$.

Figure 4 shows a contour plot of the number of stars per square degree brighter than 17th magnitude expected to show detectable microlensing.  Figure 4 is plotted in Galactic coordinates and is centered on the Galactic Center.  The plot shows that the most probable place to find gravitational lensing in our Galaxy is first toward the Galactic center, and next in our Galactic plane.  In the direction of the Galactic center, the success rate of a lensing search monitoring all stars brighter than 17th magnitude would be on order 3 x 10$^
{-4}$ lenses degree$^{-2}$.

These results estimate the {\it static} probability of gravitational lensing -- assuming stars don't move.  But stars do move and lensing events have finite durations.  The methodology for making {\it dynamic} lensing probability estimate, allowing for relative lens, observer, and source motion, was discussed previously by Nemiroff (1991).

Taking the detection volume defined above and allowing it to move will allow it to ``catch" increasingly more lenses as a function of time.  Were the source to move with transverse speed $v$ relative to the observer for a time $t$, it would sweep out a volume
 \begin{equation}
 V_{\rm dynamic} = (1/2) \pi a b v t = 
             (\pi/4) R_S^{1/2} D_{OS}^{3/2} \Phi_{app}^{1/2} v t ,
 \end{equation}
were $a = D_{OS}/2$ is the semi-major axis, and $b = \sqrt{R_S D_{OS} \Phi_{app}}$ is the semi-minor axis of the ellipsoidal detection volume.

Given canonical values of $R_S = 3$ km, $D_{OS} = 1$ kpc, $\Phi_{app} = 1/2$, $v = 100$ km sec$^{-1}$, and $t =$ 3.1 x 10$^7$ sec, the ratio of dynamic to static volumes is about 5.  This means that the probability of lensing is about 5 times as great over the course of a year than during a single observation.

In general, the duration of a microlensing events would be on order $t_{dur} \sim 2b/v$. Given the canonical values listed above, a typical duration for a microlensing event would be $t_{dur} \sim 50$ days. The brightness of the lens and the maximum amplification of the event is automatically included in $\Phi_{app}$.  In general, higher amplification events, closer sources, and brighter lenses would cause shorter duration events.  

From Figure 3, the magnitude at which $N = 1/5$ is about 15th magnitude without magnification bias, and about 14th magnitude with magnification bias.  Therefore, were all stars brighter than 15th magnitude monitored for a year with observations spaced shorter then the event duration, at least one is likely to show an $A > 1.34$ event. 

\section { Discussion and Conclusions }

Directly applying stellar densities derived from the Bahcall-Soneira model to microlensing leads to several inaccuracies.  For example, this model is not designed to directly indicate the space density of binary stars, dark matter lenses, or the Galactic bar, each of which could alter the microlensing rate.  No corrections for these were made in the present calculation, although it is believed that these factors would only increase the actual observed lensing rate.

One conclusion that can be drawn is that microlensing effects are present in {\it any} set of 100 million source stars that can possibly be monitored.  In fact, the brightest 100 million stars are among the {\it least} likely set to show microlensing, since their proximity allows relatively little room to place the lenses.  Additionally, these stars are so close that wide-angle measurements are needed just to monitor them.  Searching for all sky microlensing, however, should yield the closest lenses and the brightest sources.

The relative brightness of these microlensing events might make them more than a curiosity. At 15th magnitude, most observatories would be capable of follow-up coverage of an announced, on-going microlensing event.  This could lead to microlensing events with extraordinarily good time coverage.  Good time coverage, in turn, creates an increased chance of recovering brief bumps in microlensing light curves caused by planets orbiting the stellar lens (see, for example, Mao \& Paczynski 1991; Gould \& Loeb 1992).

The relative proximity of these lenses and/or sources might be expected to create an increased chance that parallax could be measured for them, and hence foster the ability to better understand lens and source distances, masses, and even transverse velocities.  For lenses crossing the disk of the source, the lens will act like a huge magnifying glass being swept across the source star face.  Potentially recoverable information includes the radius, shape, and brightness distribution of the source star (Nemiroff \& Wickramasinghe 1994; Gould 1994; Witt \& Mao 1994; Gaudi \& Gould 1998).

The relatively high angular separation of these microlensing events might be expected to allow optical depth estimates toward widely separated directions in our Galaxy.  Therefore, event rates would give information on the Galactic stellar and dark matter distributions.  The optical depth to lensing has yet to be accurately predicted in any direction -- the result is always a surprise.  Possibly optical depth estimates along new directions would also yield new surprises.

Some might consider the above results discouraging, indicating that the number of stars is too high, the event rate too low, the data rate too high, and the monitoring instruments needed are too unusual to search for microlensing over the entire sky.  Others might consider that the technology need to carry out a large area search capable of detecting microlensing is not only possible, but surveys that could be considered prototypical are already being implemented.
Examples of such monitoring systems include LOTIS (Park et al. 1997) and ROTSE (Marshall et al. 1997), originally created to search for optical counterparts to gamma-ray bursts.  LOTIS is already generating daily images of approximately $\pi$ steradians down to 13th magnitude while searching for the optical counterparts of gamma-ray bursts.  A planned improvement creating ``SUPER-LOTIS" will scan the sky every three weeks down to a limiting visual magnitude of 19, at which time bright star microlensing events would be expected to exist in its database.

Lastly, it should be pointed out that the ``secondary" information on the intrinsic variability of bright star systems gathered in a wide-angle microlensing search should be of significant value. Interesting causes of variability includes stellar pulsations, binary star eclipses, and recurrent novae.  Arguably, since transient events of even unknown cause could be followed by small telescopes common both to professionals and amateurs, ``triggers" created by a wide-angle monitor could be of interest regardless of whether they were related to microlensing.

I thank C. Ftaclas and J. B. Rafert for helpful comments, G. Marani for a careful reading of the manuscript, and B. Paczynski for pointing out parallax possibilities. This research was supported by grants from NASA and the NSF.

\clearpage

\figcaption[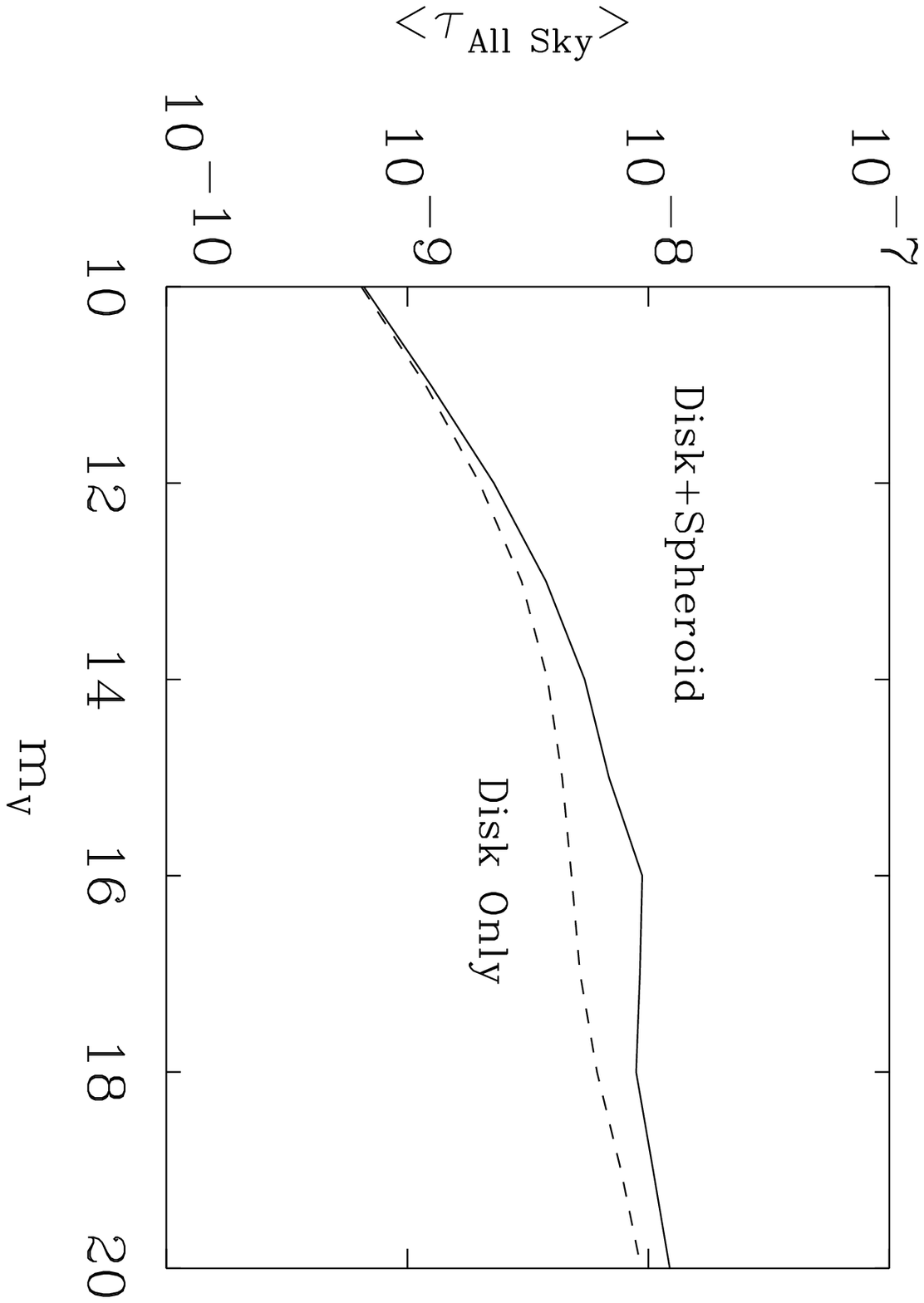]{A plot of the optical depth to gravitational microlensing averaged over the entire sky, versus apparent visual magnitude.  The solid line depicts the optical depth to a model incorporating both the Galactic disk and the central spheroid, while the dashed line depicts optical depth including only the disk.
\label{fig1}}

\figcaption[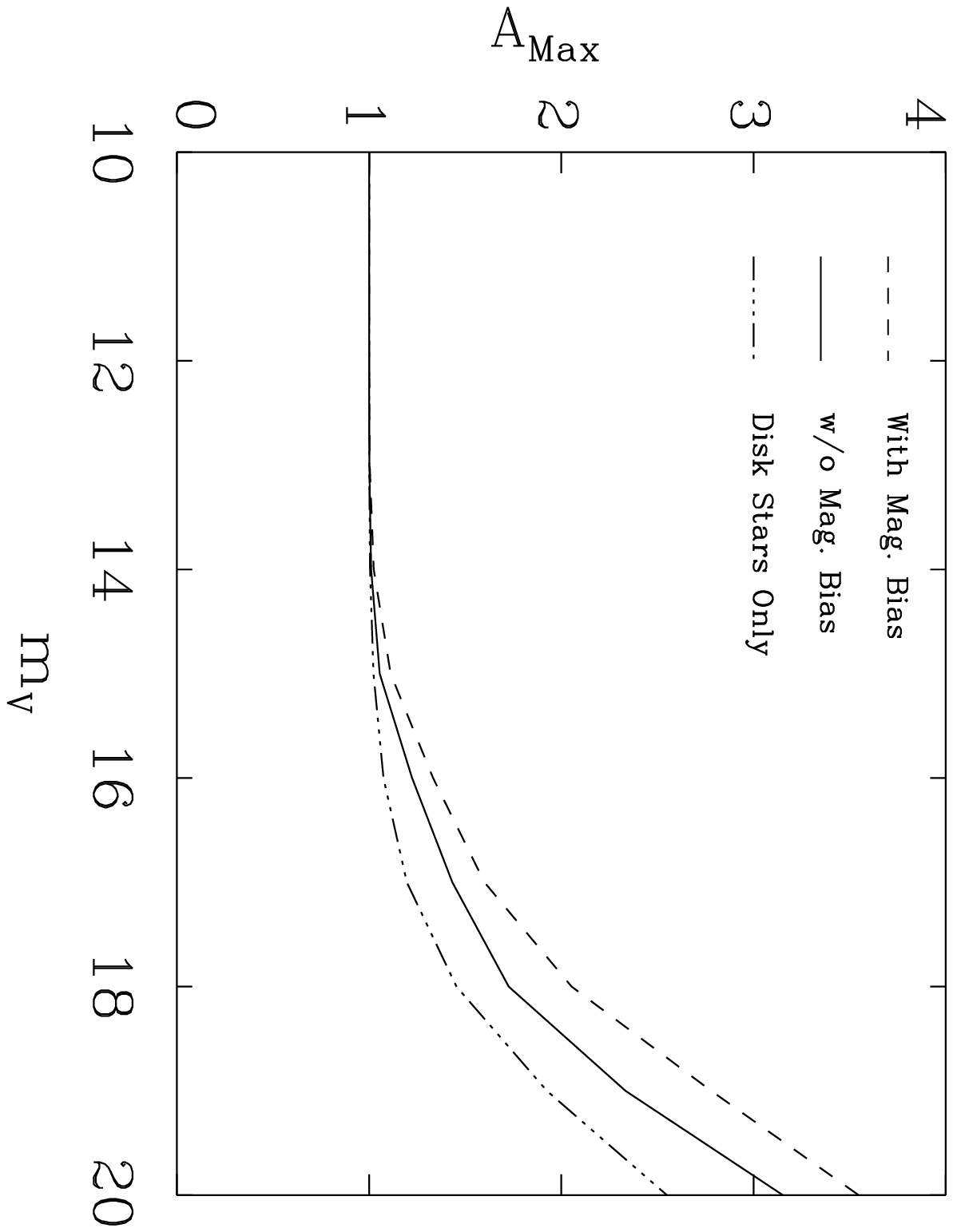]{A plot of the maximum expected gravitational-lens induced amplification expected over the entire sky, as a function of the faintest visual magnitude observable.  The solid line depicts a survey of a fixed sample of stars so that no magnification bias should operate.  The dashed line depicts a survey where stars below the survey magnitude limit could be microlensed above the survey limit: magnification bias.  The dot-dashed line depicts a survey of only disk stars without magnification bias.
\label{fig2}}

\figcaption[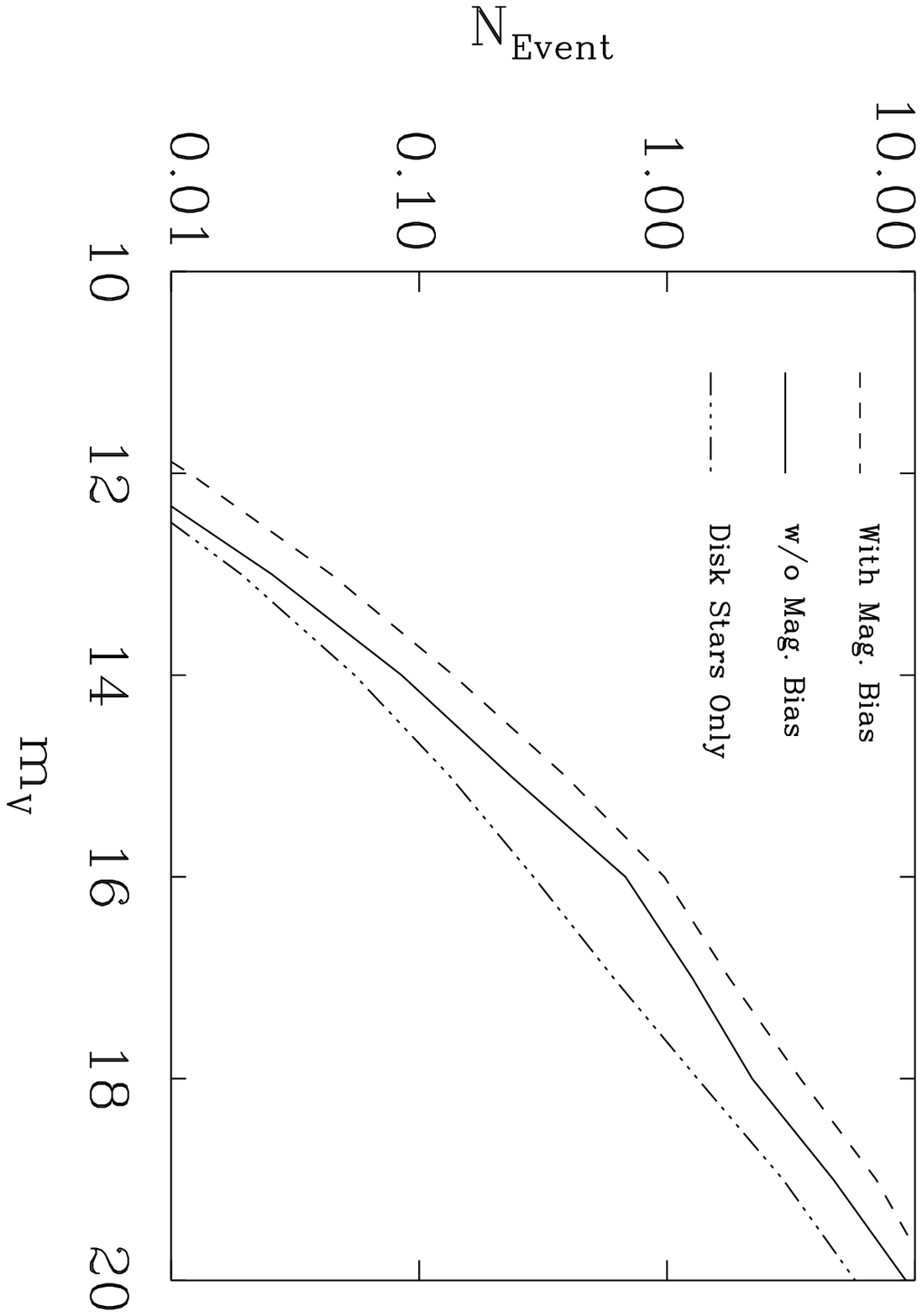]{A plot of the number of lensing events with apparent amplitude $A > 1.34$ expected over the entire sky at any time, as a function of the faintest visual magnitude observable.
\label{fig3}}

\figcaption[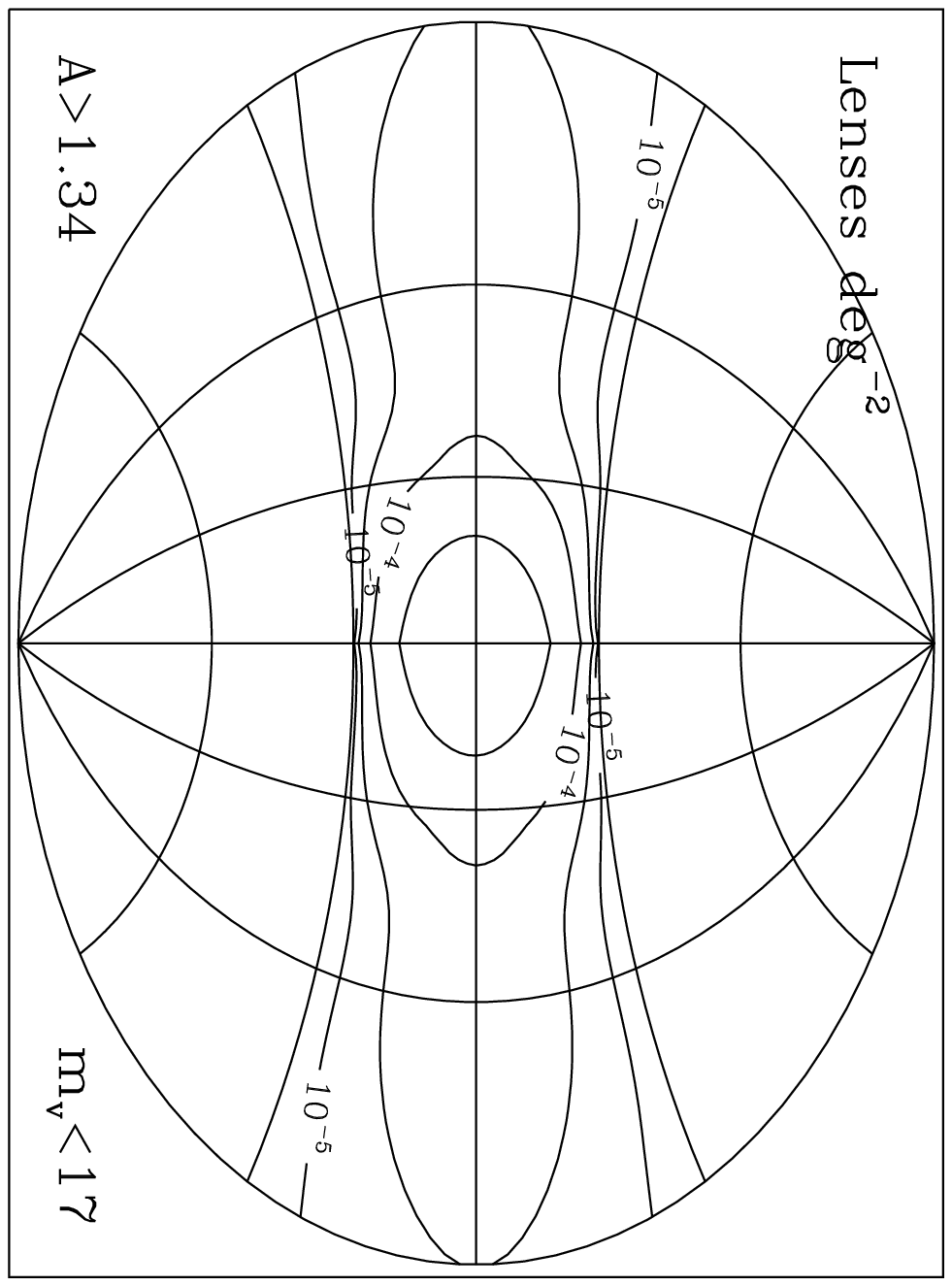]{A plot of our Galaxy contoured by the expected surface density of microlensing, in number of events with $A > 1.34$ per square degree, without magnification bias. The Galactic center is at the origin.
\label{fig4}}

\plotone{mtau.eps}

\plotone{ma.eps}

\plotone{mnlens.eps}

\plotone{tau17lb.eps}

\end{document}